\documentstyle[epsf,fleqn,12pt]{article}

\advance\textheight by \topskip
\newcommand{\be}[1]{\begin{equation} \label{#1}}
\newcommand{\ee}{\end{equation}}
\def\beqar{\begin{eqnarray*}}
\def\eeqar{\end{eqnarray*}}
\def\biggi{\bigskip\noindent}

\newcommand{\p}{\partial}
\newcommand{\dps}{\displaystyle}

\newcommand{\wt}[1]{\widetilde{#1}}

\def\R{{\rm I\! R}}

\def\C{{\rm C\kern-4.7pt
\vrule height 7.7pt width 0.5pt depth 0pt \phantom {.} \hspace{1pt} }}

\def\Im{{\rm Im}\, }
\def\Re{{\rm Re}\, }

\def\schnitt{\cap}
\def\Schnitt{\bigcap}
\def\verein{\cup}
\def\Verein{\bigcup}

\def\d{\delta}
\def\e{\epsilon}
\def\bfN{{\bf N}}
\def\bfS{{\bf S}}
\def\bfI{{\bf I}}

\def\P{P^{\psi_0}}

\def\s{\sigma}
\def\Si{\Sigma}

\def\hil{{\cal H}}
\def\dom{{\cal D}}

\def\No{{\cal N}}
\def\G{{\cal G}}
\def\Gedr{{\cal G}^{\epsilon  \delta r}}

\def\vpsi{v^\psi }
\def\vpsit{v^{\psi _t}}
\def\I{{\cal I}}

\newcommand{\wf}{wave function}
\newcommand{\BM}{Bohmian mechanics}
\newcommand{\sa}{self-adjoint}
\newcommand{\Sa}{self-adjointness}
\newcommand{\Ham}{Hamiltonian}
\newcommand{\Schr}{Schr\"{o}dinger}
\newcommand{\Sseq}{Schr\"{o}dinger's equation}
\def\diff{differentiable}

{\begin{list}{}%
{\settowidth{\labelwidth}{#1}%
\setlength{\leftmargin}{\labelwidth}%
\addtolength{\leftmargin}{\labelsep}%
\setlength{\parsep}{0pt}%
\setlength{\itemsep}{0pt}%
\setlength{\topsep}{0pt}%
}}%
{\end{list}}%

\begin{document}

\noindent
Contribution to ``Bohmian Mechanics and Quantum Theory: An Appraisal,''\\
edited by J.T.\ Cushing, A.\ Fine, and S.\ Goldstein

\bigskip

\biggi
{\huge \bf Global existence and uniqueness}

\medskip\noindent
{\huge \bf of Bohmian trajectories}

\bigskip

\biggi
Karin Berndl\\
Mathematisches Institut der Universit\"{a}t M\"{u}nchen,\\ Theresienstra{\ss}e
39, 80333 M\"{u}nchen,
Germany

\biggi
\today

\section{Introduction}

Consider the equations of motion of \BM\
for a system of $N$ particles with
masses $m_{1},...,m_{N}$ moving in physical space $\R^3$: the \wf\ $\psi$
evolves according to
\Schr 's equation
\begin{equation}  \label{seq}
i\hbar \frac{\partial \psi _t(q)}{\partial t} = \left( -\sum_{k=1}^{N}
\frac{{\hbar}^{2}}{2m_{k}}\Delta_{k} + V(q)            \right)
 \psi _t(q) , \end{equation}
and the configuration $Q=({\bf Q}_1, \dots ,
{\bf Q}_N) \in \R^{3N}$,
with ${\bf Q}_k \in \R ^3$ denoting the position of the $k$-th
particle, evolves according to Bohm's equation
\begin{equation} \label{bohmev}
\frac {dQ_t}{dt} = \vpsit (Q_t) ,
\end{equation}
where the velocity field $v^{\psi}=({\bf v}^{\psi}_{1},
\dots ,{\bf v}^{\psi}_{N})$ is determined by the \wf \ $\psi$
\be{velfield}
{\bf v}_{k}^{\psi}(q) = \frac{\hbar}{m_{k}}
\mbox {Im} \frac{{\nabla}_{k}\psi (q)}{\psi (q)}
\ee
(Bohm 1952). This velocity field is regular on the subset of $\R^{3N}$ where
$\psi\neq 0$ and is \diff . The
question arises as to what happens if the configuration $Q_t$, when moving in
accordance with (\ref{bohmev})
along  integral curves of $\vpsit$, reaches at some time $\tau$ a singularity
of $\vpsi$, for example a node of the
\wf , i.e.\ a point where $\psi =0$. In general, the event of reaching a
singularity of $\vpsi$ corresponds to a
singularity of the motion: the velocity $dQ_t/dt$ becoming infinite, $Q_t$
being discontinuous, or even
``exploding,'' i.e.\ reaching infinity as $t\to \tau$.
In some cases it may be possible to continue the Bohmian trajectory through a
singularity of the velocity field,
but in any case, Eqn.\ (\ref{bohmev}) is not defined at singularities of
$\vpsi$, and the theory would have to be
supplemented by a suitable prescription for how to extend the motion through a
singularity, or which trajectories
to avoid.

We should therefore consider the following problem---the {\em initial value
problem in \BM : For given initial
values $\psi_0$ and $Q_0$ at a time $t_0$ (we shall put $t_0=0$), do there
exist global unique solutions
$(\psi_t, Q_t)$ of (\ref{seq},\ref{bohmev}) such that $\psi_{t_0}=\psi_0$ and
$Q_{t_0}=Q_0$?} A positive
answer to this question for all or at least for a suitable majority of initial
conditions, or for a class of ``physically
relevant'' initial conditions, is certainly important: the limits of global
existence of solutions may hint at limits of
validity of the theory.

\begin{figure}[p]
\begin{center}
\leavevmode
\epsfysize=10cm
\epsffile{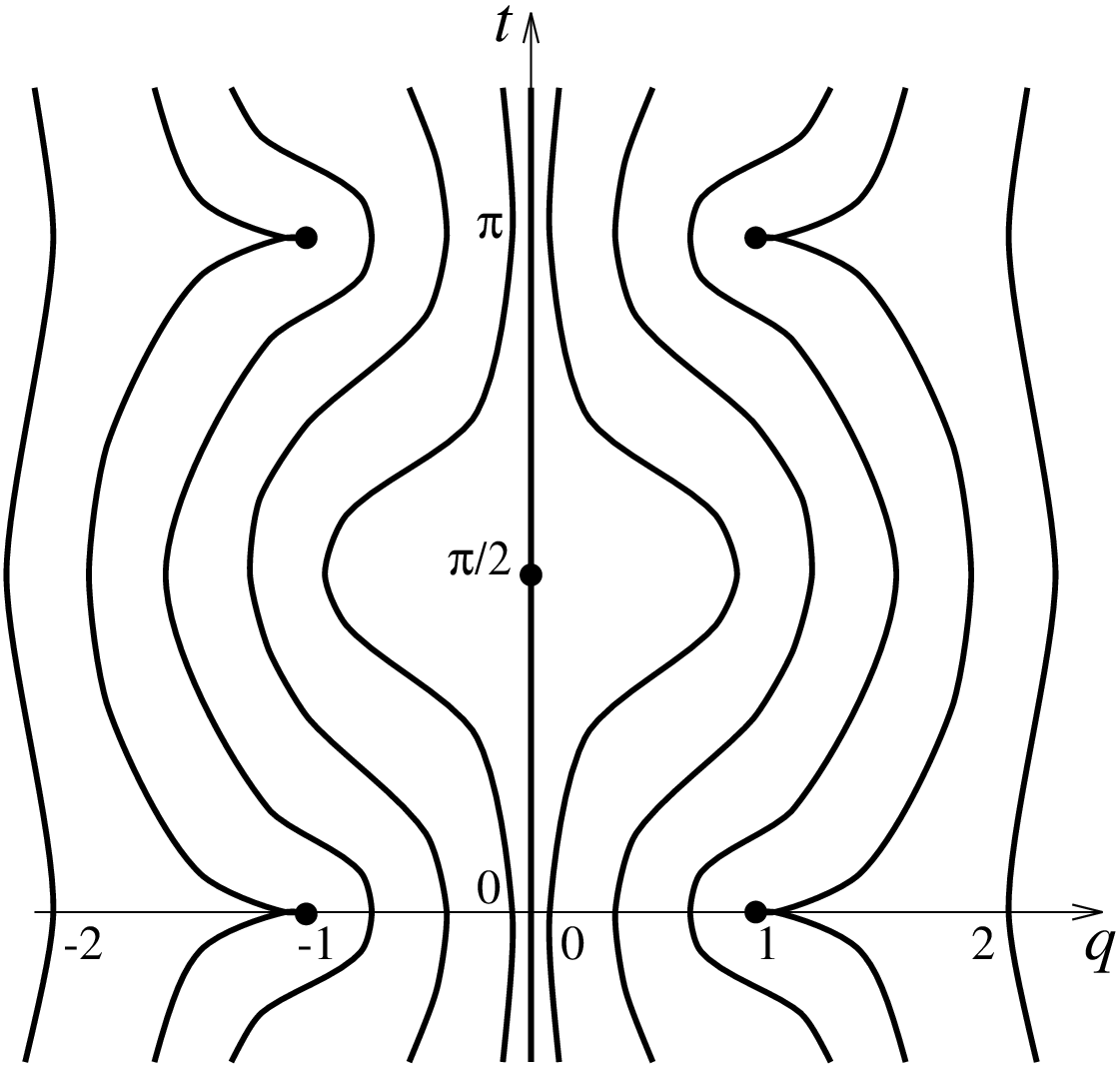}
\end{center}
\caption{Sketch of the course of Bohmian trajectories for the \wf \ (4). Nodes
of the \wf\ are marked by dots.}
\label{nodes}
\end{figure}

Let us first give a simple explicit example where indeed some trajectories
reach nodes of the \wf . Consider the
one-dimensional harmonic oscillator (with $\hbar = m=\omega =1$), i.e, the
potential $\dps V(q)=q^2/2$, and
take as the \wf\ a superposition of the ground state and the second excited
state:
\be{exwf}
\psi_t(q)=e^{-q^2/2}\Bigl( 1+(1-2q^2)e^{-2it}\Bigr) e^{-it/2}.
\ee
This \wf\ leads to a velocity field $\vpsi$ which is an odd function of $q$ and
periodic in time with period $\pi$.
The Bohmian motion is therefore invariant under reflection $q\to -q$ and
periodic in $t$ with period $\pi$: If
$Q(t)$ is a solution of (\ref{bohmev}), then $-Q(t)$ is also a solution of
(\ref{bohmev}), and $Q(t+\pi)=Q(t)$.
The \wf\ $\psi$ has nodes at $(q,t)=(0, (n+ 1/2)\pi )$ and at $(q,t)= (\pm 1,
n\pi )$ for all integers $n$. There are
three trajectories which periodically run into nodes of $\psi$, see Figure
\ref{nodes}. One of them, the constant
$Q_t=0$ for $t\neq (n+1/2)\pi$, is a solution of (\ref{bohmev}) which runs
with velocity 0 at times  $t\to  (n+
1/2)\pi$ into nodes of $\psi$.
The  other two ``node-crossing'' trajectories (which are reflection images of
each other) are singular at the nodes:
for example, in the vicinity of the node at $q=1$, $t=0$ the trajectory running
into it has the form $\dps Q_t=
(3t^2 /4)^{1/3} +1$, i.e.\ it has infinite velocity at $t=0$.

In this example the trajectories may be continued through the nodes in an
obvious and consistent way. This is an
artifact of the low dimensionality: In fact, in $d=1$ dimension, $Q_t$
satisfies
\be{smap} \int_{-\infty}^{Q_{t}(q_0)} |\psi_t(q)|^{2} \, dq
=\int_{-\infty}^{q_0}
|\psi_0(q)|^{2} \, dq, \ee
employing the equivariance of the $|\psi_t|^2$-measure, i.e.\ that the density
$\rho_0=|\psi_0|^2$ evolves under
the Bohmian dynamics to $\rho_t=|\psi_t|^2$ (D\"urr, Goldstein, Zangh\`\i\
1992a; see also Section 4), and the
property that trajectories do not cross in configuration-space-time.
Clearly, much less regularity of $\psi$ is needed for this definition of the
motion of the Bohmian
configuration---in particular, nodes of the \wf\ are no problem---and, if both
definitions (\ref{bohmev})
and (\ref{smap}) are
possible, they agree.
However, a generalization of (\ref{smap}) to the physically most interesting
case of $d=3N$-dimensional
configuration space is not known. We shall turn now to the question of the
existence of global unique solutions of
the equations of motion of \BM , (\ref{seq}) and (\ref{bohmev}).

\section{The Schr\"odinger equation}

In Bohmian mechanics, the evolution of the \wf\ $\psi$ according to
Schr\"odinger's equation (\ref{seq}) is
independent of the evolution of the actual particle configuration $Q$ according
to the guiding equation
(\ref{bohmev}), while for
the integration of (\ref{bohmev}) we need $\psi$. Therefore, when solving the
initial value problem, we may consider Schr\"odinger's equation first.

The linear partial differential equation (\ref{seq}) is usually discussed in
the framework of  the Hilbert space
$\hil = L^2 (\R^{3N})$ of square integrable functions, by viewing the
Hamiltonian $\dps \wt H=-
\sum_{k=1}^{N}
\frac{{\hbar}^{2}}{2m_{k}}\Delta_{k} + V(q) $ as the generator of the unitary
group $U_t=e^{-iHt/\hbar }$
(Kato 1951, Reed and Simon 1975).
Usually, boundary conditions on the \wf\ have to be specified in order to get a
unique time
evolution $(U_t)_{t\in\R}$. In more technical terms, one has to select a
self-adjoint extension $H$ of the partial
differential operator $\wt H$, which is a priori defined only on sufficiently
smooth functions. As different
self-adjoint extensions generate different time evolutions of the \wf , the
choice of the right self-adjoint
extension is a
matter of physics.
We shall comment on this in Section 4.
Given $(U_t)_{t\in\R}$, for any
initial $\psi_0\in\hil$ we have a global unique \wf\ $\psi_t=U_t\psi_0$. This
\wf , however, in general is not a
genuine solution of \Sseq : Generic $\psi_0\in\hil$ are not smooth functions,
and $\psi_t$ will not be \diff , so
that the differential equation (\ref{seq}) cannot be discussed. The standard
procedure now is to forget about
\Sseq , and argue that all one is interested in is a unitary time
evolution---or, even more abstractly, a unitary
representation of the time translation group. This is, however, not sufficient
for our approach. {From} the point
of view of \BM , the \wf\ $\psi$ is a smooth field on configuration-space-time
solving \Sseq . To obtain this, we
have to put suitable conditions on the initial \wf\ $\psi_0$. It turns out that
a suitable subspace of $\hil$ is given
by
the so-called $C^\infty$-vectors of the Hamiltonian $H$, $\dps C^\infty (H)=
\Schnitt_{n=1}^\infty \dom (H^n)$. $\dom (H^n)$ denotes the domain of the
$n$-th power of the Hamiltonian,
i.e. the set of \wf s  for which the expectation value of the $2n$-th power of
$H$, the ``energy,'' is finite, $\dps
\int_{\R^{3N}} (H^n\psi )^2\, dq= \langle \psi | H^{2n}\psi \rangle
<\infty$.$^1$ $C^\infty (H)$ is a dense
subset of $\hil$, and invariant under the time evolution $U_t$. Eigenfunctions
and wave packets are special
$C^\infty$-vectors, so all \wf s usually considered in physics are included. It
should not at all be regarded as a
defect of \BM\ that it is not defined for all $\psi_0\in\hil$: {From} the point
of view of \BM , the Hilbert space
$L^2(\R^{3N})$ is not the state space of the \wf , but a useful tool for the
analysis of the theory. (In this context,
see also the contributions of D\"urr, Goldstein, Zangh\`\i\ and Daumer to this
volume.)

We have: {\em For any $\psi_0\in C^\infty (H)$, $\psi_t= U_t\psi_0=
e^{-itH/\hbar}\psi_0$ is a global smooth
solution of \Sseq \ on}\/ $\Omega\times \R$, where $\Omega$ denotes the set
where the potential is smooth.
This result cannot be genuinely strengthened: the \wf\ cannot be expected to be
regular at points where the
potential is singular. For instance, the ground state eigenfunction of the
Coulomb potential $V(q)=1/|q|$, $e^{-
|q|}$, is not \diff\ at the singularity of the potential at $q=0$. Important
examples of potentials are the
$N$-particle Coulomb interaction for particles with charges $e_i$
\[ V_{\rm Coulomb}({\bf q}_1,\dots ,{\bf q}_N) = \sum _{i=1}^N \sum _{j=i+1}^N
\frac{e_ie_j}{|{\bf q}_i-
{\bf q}_j|} \]
with $\dps  \Omega_{\rm Coulomb}= \R^{3N} \setminus \Verein _{i=1}^N \Verein
_{j=i+1}^N \{ {\bf
q}_i={\bf q}_j\} $,
and the $N$-electron atom
\[ V_{\rm atom}({\bf q}_1,\dots ,{\bf q}_N) =\sum _{i=1}^N  \frac{e_ne_i}{|{\bf
q}_i|} + \sum _{i=1}^N \sum
_{j=i+1}^N \frac{e_ie_j}{|{\bf q}_i-{\bf q}_j|} \]
with $\dps \Omega_{\rm atom}= \R^{3N} \setminus \Biggl( \Verein _{i=1}^N \{
{\bf q}_i=0\} \verein \Verein
_{i=1}^N \Verein _{j=i+1}^N \{ {\bf q}_i={\bf q}_j\} \Biggr) ,$
in the approximation that the nucleus with charge $e_n$ is at rest at the
origin, acting like an external Coulomb
field.

\section{The Bohmian trajectories}

With the \wf\ $\psi\in C^\infty (\Omega\times \R)$, the velocity field $\vpsi$
(\ref{velfield}) can be formed on
the set where the \wf\ $\psi\neq 0$. We shall call the set where the velocity
field is regular $\G = (\Omega\times
\R)\setminus \No$, where $\No$ is the (space-time) set of nodes of the \wf\
$\No = \{ (q,t)\in\Omega\times\R : \psi_t(q)=0\}$.
On this set, the first order ordinary differential equation (\ref{bohmev}) is
locally integrable. If we extend the
solution as far as possible, we obtain for each initial value $q_0\in\G_0$,
i.e.\ $(q_0,0)\in\G$, a {\em maximal
solution} $Q(t; q_0)$ on a maximal time interval of existence $(\tau^- (q_0),
\tau^+ (q_0))$. The solution is
called {\em global}\/ if $\tau^-=-\infty$ and $\tau^+= +\infty$. In the
introduction we have given an example
showing that we cannot expect to have global solutions for {\it all}\/ initial
values $q_0$. But what indeed holds
true is that for {\em almost all}\/ initial configurations we have global
unique solutions. ``Almost all'' here is with
respect to the natural---the equivariant---measure $P^{\psi_0}$ on
configuration space $\R^{3N}$, that is the
measure with the density $|\psi_0|^2$.
We have proved that {\it for a large class of potentials, including the
N-particle Coulomb interaction with
arbitrary masses and charges, as well as arbitrary positive potentials,
$P^{\psi_0} ( \tau ^+ = +\infty $ and $\tau^-
=-\infty )=1$}.
In other words, for almost any initial point $Q_0$, the solution of the guiding
equation (\ref{bohmev}) is global,
i.e.\ nodes of the \wf\ or other singularities
of the velocity field will not be reached in finite time, and the solution does
not ``explode,'' i.e.\ reach infinity in
finite time.
(Cf.\ the example of Figure \ref{nodes}: the set of initial configurations for
which the solution of (\ref{bohmev})
is not global consists of 3 points---certainly a set of $P^{\psi_0}$-measure
0---while for all the other initial
values the solution is global.)

We shall comment on the proof---which is in fact quite intuitive---in the next
section. (For the details, see Berndl,
D\"urr, Goldstein, Peruzzi, Zangh\`\i\ 1995.)  Despite the relative simplicity
of the proof, the generality of the
result is rather surprising. The analogous problem in Newtonian
mechanics---global existence and uniqueness of
solutions of the $N$-body problem for Newtonian gravity---is a classical
problem of mathematical physics which
has been investigated with a great variety of  methods (Moser 1973, Diacu
1992), but for which the analogous
result---global existence of solutions for Lebesgue-almost all initial values
in the $N$-particle phase space
$\R^{6N}$---has not yet been established. In addition to the possibility of
collision singularities, the $N$-body
problem with $N>3$ yields marvelous scenarios of so-called pseudocollisions,
where some particles, while
oscillating wildly, reach infinity in finite time. Examples of such
catastrophies have been constructed by Mather
and McGehee 1975,$^2$ by Gerver 1991, and by Xia 1992.
For special situations, such as for example an almost planar solar system with
weakly eccentric planets, the KAM
theorem furnishes among other things the existence of global solutions for a
set of initial values of positive
Lebesgue measure in phase space (cf.\ for example Arnold 1963).  For the
general case of the $N$-body problem
with $N\geq 5$ it is still open whether for almost all initial values the
solutions of Newtonian gravity are global
or whether pseudocollisions occur on a set of positive measure.

The explanatory power of Newtonian mechanics rests largely on the analysis of
specific solutions of the
equations of motion, such as for example the planetary motion, or the gyroscope
motion---physical situations
where the bodies are modelled by fixed mass densities. The more ambitious
program of Newtonian mechanics for
a system composed of a huge number of point particles or rigid balls, for
example the statistical gas theory, is
difficult for two reasons. Firstly, to get global existence of solutions for
almost all initial values---certainly a
prerequisite of a statistical analysis---one has to deal for example with the
problem of singular  interactions in the
case of point particles resp.\ multiple collisions in the case of rigid balls.
Secondly, more interesting than
equilibrium properties and much more difficult to analyse are the physical
effects occuring in the transition
{}from nonequilibrium to equilibrium.

In \BM , actual trajectories are interesting in some cases. However, to
determine the equilibrium properties they
are not needed. And in contrast to the situation in classical physics, we have
the strongest empirical evidence that
our world is in quantum equilibrium (D\"urr, Goldstein, Zangh\`\i\ 1992a and
1992b). Our theorem of global
existence of Bohmian trajectories is thus exactly what is necessary and
sufficient for a complete analysis of \BM .
It provides the rigorous basis for the derivation of the quantum formalism as
well as  scattering theory {from} an
equilibrium analysis of \BM .  (These topics are described in the contributions
of D\"urr, Goldstein, Zangh\`\i\
and Daumer to this volume.)
In this sense, what we have proved is just that $|\psi |^2$ {\it is an
equivariant density for the Bohmian particle
motion,}
\be{equivar} \rho _0=|\psi _0|^2 \quad \Longrightarrow \quad \rho _t=|\psi
_t|^2 \quad \mbox{ for all } t\in\R. \ee

\section{The quantum flux}

The preceding sentence may have confused readers: Why isn't the equivariance of
the $|\psi _t|^2$-measure clear
without an intricate examination of the global existence and uniqueness of
solutions of \BM ? Why doesn't
(\ref{equivar}) follow immediately {from} a comparison of the continuity
equation for an ensemble of
configurations moving with velocity $\vpsi$ and having a density $\rho _t(q)$
\begin{equation}
\frac{\partial}{\partial t} \, \rho _t(q) \ + \ \sum_{k=1}^N \mbox{div}_{k}\,
\bigl( {\bf v}^{\psi _t}_{k}(q) \, \rho _t(q)\bigr) \  = \ 0  \label{conteq}
\end{equation}
with the ``quantum continuity equation''
\begin{equation}
\frac{\partial}{\partial t} \,  |\psi_t(q)|^{2} \ + \ \sum_{k=1}^{N}
\mbox{div}_{k} \ {\bf j}^{\psi_t}_{k}(q)  \ =\ 0 , \label{qfluxeq}
\end{equation}
noting that the quantum probability current
 $j^{\psi}=({\bf j}_{1}^{\psi},...,{\bf j}_{N}^{\psi})$ is given by
\[ {\bf j}_{k}^{\psi} \ = \ {\bf v}_{k}^{\psi}\, |\psi|^{2} \ =\
\frac{\hbar}{{m}_{k}} \, \Im ({\psi}^{\ast}  {\nabla}_{k} \psi ) \ ?  \]
The answer is that the continuity equation (\ref{conteq}) expresses how an
ensemble density $\rho_0$ evolves
under the deterministic evolution of the trajectories, and holds therefore on
just that set which is covered by the
integral curves of $\vpsit$. We shall denote this set---the image set of the
maximal solution $Q$---by $\I$:
\[ \I = \{ (q,t)\in\G : \exists q_0\in\G_0\ \mbox{with}\ t\in
(\tau^-(q_0),\tau^+(q_0))\ \mbox{and}\ q=Q_t(q_0)\} .
\]
Equation (\ref{qfluxeq}), on the other hand, is an identity for every $\psi_t$
which solves \Sseq . Thus
(\ref{equivar}) holds
on $\I$: $\rho_t(q)= |\psi_t(q)|^2$ for $(q,t)\in\I$. By putting $\rho_t(q)=0$
on the set which is not reached by the
Bohmian trajectories $(\Omega\times\R)\setminus \I$, we obtain $\rho_t(q)\leq
|\psi_t(q)|^2$ for $(q,t)\in
\Omega\times\R$.
This insight is fundamental to the proof of global existence.

The second fundamental insight comes from the consideration of the
configuration-space-time flux $J(q,t) =
(\vpsit (q) \rho_t(q) , \rho_t(q))$. To establish $P^{\psi_0} ( \tau ^+ =
+\infty $ and $\tau^-=-\infty )=1$, i.e.\ that
for $\P$-almost all initial configurations $q_0$ the maximal solution is
global, we have to show that the solution
will reach singularities of the velocity field or infinity in finite time at
most for a set of initial configurations of
$\P$-measure zero.
The probability of such a ``bad event'' is estimated with the help of the flux
$J(q,t)$:
For any hypersurface $\Si$ in configuration-space-time with local normal vector
field $n(q,t)$ and surface area
element $d\s$,
\[ \int _\Sigma |J(q,t)\cdot n(q,t)| \, d\sigma\]
is the expected number of crossings of the hypersurface $\Sigma$ by the Bohmian
trajectory and hence a bound
for the probability of crossing $\Sigma$.
We obtain
\be{probest} \P(Q_t\mbox{ crosses }\Si )\leq \int _\Sigma |J(q,t)\cdot n(q,t)|
\, d\sigma \leq \int _\Sigma
|J^{\psi_t} (q)\cdot n(q,t)| \, d\sigma \ee
with the quantum flux $J^{\psi_t} (q) = (j^{\psi_t}(q), |\psi_t(q)|^2 )$.

\begin{figure}[p]
\begin{center}
\leavevmode
\epsfysize=10cm
\epsffile{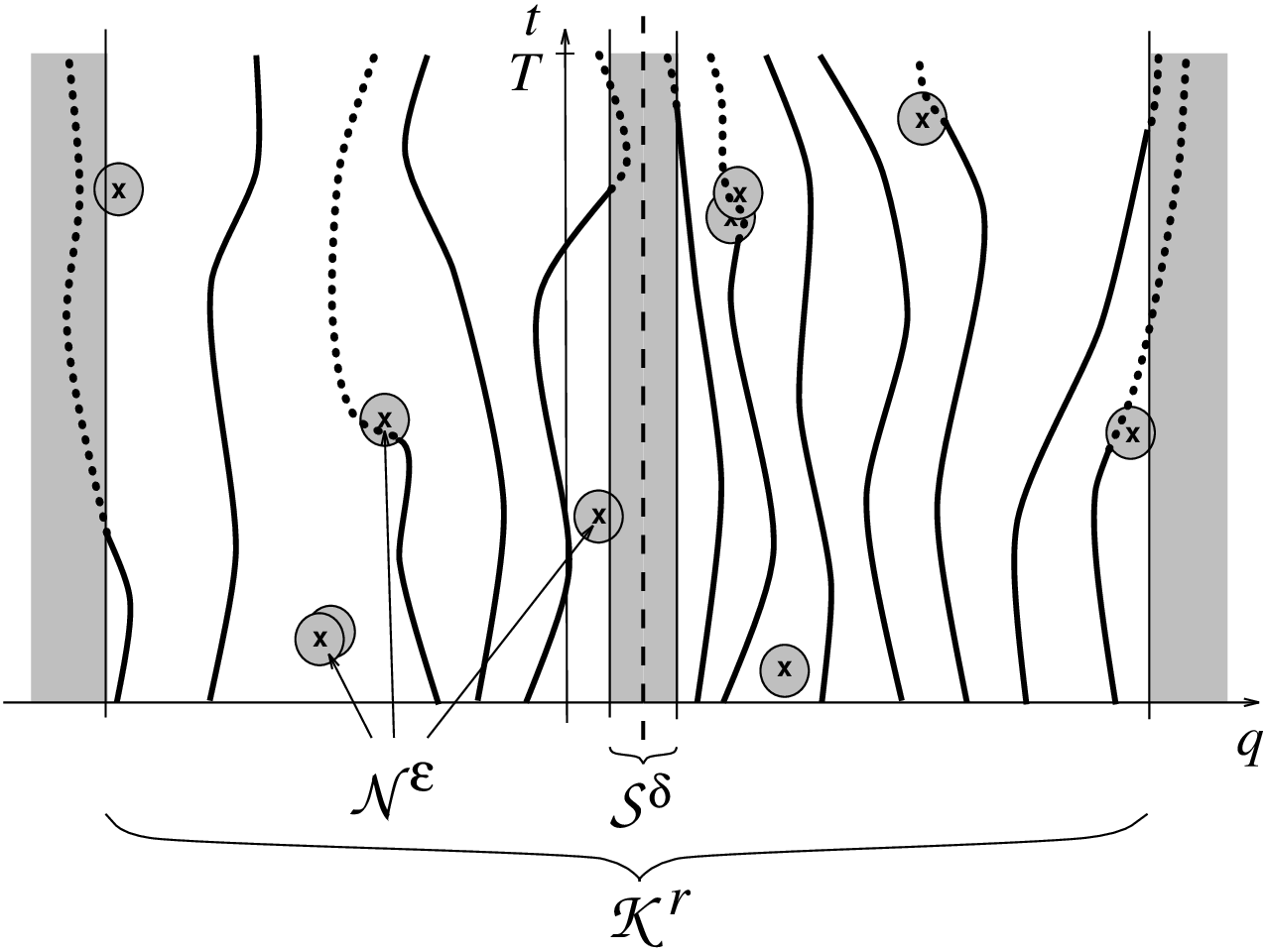}
\end{center}
\caption{Bohmian trajectories run in the white area $\Gedr = (({\cal K}
^r\setminus {\cal S} ^\delta )
 \times \R ) \setminus {\cal N} ^\epsilon $. The set of singularities of the
potential  ${\cal S} \times \R$ is dashed,
nodes of the \wf\ are marked by {\sf x}. Trajectories having crossed $\p\Gedr$
are dotted. By letting the grey
neighborhoods ${\cal N}^\e$ and ${\cal S}^\d$ around the singularities of the
velocity field shrink and the
sphere ${\cal K}^r$ blow up, the set of dotted trajectories shrinks to a set of
measure zero.}
\label{gednbild}
\end{figure}

Consider now neighborhoods around the singularities of the velocity field:
${\cal N}^\epsilon $, a
(configuration-space-time) neighborhood of thickness $\epsilon$ around the set
of nodes ${\cal N}$
of the \wf , ${\cal S}^\delta$,  a (configuration space)
neighborhood of thickness $\delta$ around the set of singularities of the
potential ${\cal S} = \partial \Omega$,
and ${\cal K}^{r}$, a sphere in configuration space of radius $r$ to control
escape to infinity.  ${\cal G}
^{\epsilon \delta r}$ denotes the set
of ``$\epsilon$-$\delta$-$r$-good'' points in configuration-space-time: $ {\cal
G} ^{\epsilon \delta r}=
(({\cal K} ^r \setminus {\cal S} ^\delta )
 \times \R ) \setminus {\cal N} ^\epsilon $ (see Figure \ref{gednbild}).

A Bohmian trajectory approaching a singularity of the velocity field or
infinity first has to cross the boundary of
$\Gedr$. {From} (\ref{probest}), we obtain the following bound for
the probability of a ``bad event'' in the time interval $[0,T]$: for all
$0<T,\e ,\d ,r<\infty$
\begin{eqnarray} \P(\tau^+<T) & \leq & \P(\G_0\setminus \Gedr_0) \ +\
\int_{\p\Gedr\schnitt
(\R^{3N}\times[0,T])}
|J^{\psi_t} (q)\cdot n(q,t)| \, d\sigma \nonumber\\
& \leq & \P(\G_0\setminus \Gedr_0) \ +\ \bfN (\e,\d,r)\ + \ \bfS(\d)\ +\ \bfI
(r) \label{bound}\end{eqnarray}
with
\[ \begin{array}{l}
\dps \bfN (\e,\d,r)\ =\ \int_{\p \No^\e\schnitt (({\cal K}^r\setminus{\cal
S}^\d)\times [0,T])}|J^{\psi_t} (q)\cdot
n(q,t)| \, d\sigma ,\\
\dps   \bfS (\d)\ =\ \int_{(\p {\cal S}^\d\schnitt
\Omega)\times[0,T]}|J^{\psi_t} (q)\cdot n(q,t)| \, d\sigma ,\\
\dps   \bfI (r)\ =\ \int_{(\p {\cal K}^r \schnitt
\Omega)\times[0,T]}|J^{\psi_t} (q)\cdot n(q,t)| \, d\sigma .
\end{array} \]
By proving that for appropriate choices of sequences $\e\to 0$, $\d\to 0$, and
$r\to\infty$ the right hand side of
(\ref{bound}) gets arbitrarily small we show that for all $T>0$,
$\P(\tau^+<T)=0$, and thus
$\P(\tau^+<\infty)=0$. {From} time reversal invariance we obtain that also
$\P(\tau^->-\infty)=0$, and that
altogether the solutions of Bohm's equation (\ref{bohmev}) with the velocity
field (\ref{velfield}) are global for
almost all initial configurations.

Heuristically, it is rather immediate that the flux integrals $\bfN$, $\bfS$,
and $\bfI$ get arbitrarily small as $\e\to
0$, $\d\to 0$, and $r\to\infty$. For $\bfN$, observe that the flux $J^\psi=0$
at the nodes of the \wf . {From} the
continuity of $\psi$, $J^\psi$ is small in the vicinity of $\No$. Furthermore,
the nodal set $\No$ itself is
generically small: since $\psi$ is a complex function, the set where
$\psi_t(q)=0$, i.e.\ $\Re \psi_t(q)=0$ and
$\Im \psi_t(q)=0$, is generically of codimension 2 in configuration-space-time.
Thus the area of $\p\No^\e$
should be small.
Of the set $\cal S$ of singularities of the potential $V$, we assume in Berndl,
D\"urr, Goldstein, Peruzzi,
Zangh\`\i\ 1995 that it be contained in a union of a finite number of
$(3N-3)$-dimensional hyperplanes, as is
certainly the case for the $N$-particle Coulomb interaction $V=V_{\rm Coulomb}$
and the $N$-electron atom
$V=V_{\rm atom}$. The area of $\p{\cal S}^\d$ is therefore very small.
Moreover, in certain cases it is required
as a boundary condition for \Sa\ of the \Ham\ that
the flux into the singularities of $V$ vanishes, see below.
That the flux to infinity is small can be derived from the fact that the
quantum flux $J^{\psi_t} (q)$ tends rapidly
to 0 as $|q|\to \infty$, which follows from the square integrability of $\psi$
and $\nabla\psi$, i.e.\ from the
normalizability of the $|\psi|^2$-distribution and finite ``kinetic energy.''
These conditions are automatically
fulfilled for the considered class of potentials  for $\psi_0\in C^\infty (H)$,
which is just what we required for the
existence of a global classical solution of \Sseq .

We are thus lead back to the question of the existence of global classical
solutions of \Sseq ! In fact, the
connection between the global existence of Bohmian trajectories and global
solutions of \Sseq\ is most
remarkable. The clue is the quantum flux $J^\psi$, which in \BM\ has the
interpretation of a flux of particles
moving along deterministic trajectories with velocity $\vpsi$:
The condition that there is no flux into the critical points ensures firstly,
as explained above, that the Bohmian
configuration will not reach the critical points and thus exists globally.
Secondly, it provides suitable boundary
conditions for the domain of the \Ham\ $\dom (H)$ such that the \Ham\ will be
\sa\ on $\dom (H)$ and thus
\Sseq\ has global unique solutions as explained in Section 2. This connection
is realized by considering
that from integration by parts
\[ \int_{M} \psi^{\ast} (H\psi ) \, dq - \int_{M} (H \psi^{\ast} ) \psi \, dq
\;=\;-i\hbar \int_{\partial M} j^{\psi}\cdot n\,  ds , \]
for $M= {\cal K}^r \backslash {\cal S}^{\delta}$, and thus by the
self-adjointness of the \Ham\
\[ \lim_{\delta\rightarrow 0, r\rightarrow \infty} \left(
 \int_{\partial {\cal S}^{\delta} \cap {\cal K}^r }
j^{\psi}\cdot n\, ds \, \, + \,\,
\int_{\partial {\cal K}^{r} \setminus {\cal S}^\delta }
j^{\psi}\cdot n\,  ds \right) \ =\ 0. \]
This is only slightly weaker than the vanishing of ${\bf S}(\d)$ and ${\bf
I}(r)$ in the limit $\d\to0$, $r\to\infty$,
which is part of our sufficient condition for global existence of Bohmian
trajectories.
Moreover, in situations where the self-adjoint extension of $\wt H$ (cf.\
Section 2) is not unique, the particle
picture of \BM\ supplies an interpretation of the different possible boundary
conditions yielding different time
evolutions of the \wf\ and thus a basis for the choice of one over the others.
(For more on these points, see Berndl, D\"urr, Goldstein, Peruzzi, Zangh\`\i\
1995.)

In this way, the point of view of \BM\ provides genuine understanding to the
mathematics around the
self-adjointness of Schr\"odinger Hamiltonians!
Usually, the \Sa\ of the \Ham ---via its equivalence to the existence of a
unitary group---is motivated by the
conservation of $|\psi|^2$-probability. Probability of what? The standard
answer---the probability of {\em
finding}\/ a particle in a certain region---is justified by \BM : A particle is
found in a certain region because, in
fact, it's there. By incorporating the positions of the particles into the
theory, and thus by interpreting $|\psi|^2$ as
a probability density of particles being and the quantum flux $J^\psi$ as a
flux of particles moving, \BM\ can be
regarded as providing the foundation for all intuitive reasoning in quantum
mechanics.

\newpage
\section*{Footnotes}
\setlength{\parindent}{0pt}
\parskip 3ex plus 0.1ex minus 0.1ex

1 $\quad$ If only the expectation of $H^2$, the squared energy, is finite,
$\psi_0\in\dom (H)$, ``\Sseq\ holds in
the $L^2$-sense,'' i.e. $\dps i \lim_{h\to0}\frac{\psi_{t+h} - \psi_t}{h} =
H\psi_t$, where the equality and
convergence of the limit holds with respect to the Hilbert space norm and not
pointwise as expressed by \Sseq .

2 $\quad$ In this (one-dimensional) example the system explodes only after
infinitely many two particle
collisions; thus it does not describe a  genuine pseudocollision.

\newpage
\section*{References}

Arnol'd, V.I. (1963), ``Small denominators and problems of stability of motion
in classical and celestial
mechanics,'' {\it Russian Mathematical Surveys} {\bf 18}(6), 85--191. (Also:
Arnol'd, V.I. (1989)
 {\it Mathematical methods in classical mechanics.} New York, Springer.)

Berndl, K., D\"urr, D., Goldstein, S., Peruzzi, G., Zangh\`\i , N. (1995),
``On the global existence of Bohmian mechanics,'' to appear in {\it
Communications in Mathematical  Physics}

Bohm, D. (1952), ``A suggested interpretation of the
quantum theory in terms of ``hidden'' variables I, II,'' {\it Physical Review}
{\bf 85}, 166--179, 180--193.

Diacu, F.N. (1992),  {\it Singularities of the N-body problem}. Montr\'eal,
Les Publications CRM.
(Also: Diacu, F.N. (1993),  ``Painlev\'e's Conjecture,'' {\it Mathematical
Intelligencer} {\bf 15}, 6--12.)

D\"{u}rr, D., Goldstein, S., Zangh\`{\i}, N.  (1992a), ``Quantum equilibrium
and the origin of absolute
uncertainty,'' {\it Journal of Statistical Physics} {\bf 67}, 843--907.

D\"{u}rr, D., Goldstein, S., Zangh\`{\i}, N.  (1992b),  ``Quantum mechanics,
randomness, and deterministic
reality,'' {\it Physics Letters A} {\bf 172}, 6--12.

Gerver, J.L. (1991), ``The existence of pseudocollisions in the plane,'' {\it
Journal of Differential Equations} {\bf
89}, 1--68.

Kato, T. (1951), ``Fundamental properties of  Hamiltonian operators of \Schr\
type,'' {\it Transactions of the
American Mathematical Society} {\bf 70}, 195--211.

Mather, J., McGehee, R. (1975), ``Solutions of the collinear four body problem
which become unbounded in
finite time'' in J. Moser (ed.), {\it Dynamical Systems: Theory and
Applications}. Berlin, Springer.

Moser, J. (1973), {\it Stable and Random Motions in Dynamical Systems}.
Princeton, Princeton University
Press.

Reed, M., Simon, B. (1975), {\it Methods of
Modern Mathematical Physics II}.  San Diego, Academic Press.
(Also: Simon, B. (1977), ``An introduction to the self-adjointness and spectral
analysis of \Schr\ operators'' in W.
Thirring, P. Urban (eds.), {\it The \Schr\ Equation}.  Wien, Springer, pp.
19--42.)

Xia, Z.  (1992), ``The existence of noncollision singularities in Newtonian
systems,'' {\it Annals of Mathematics}
{\bf 135}, 411--468.

\end{document}